\documentclass{article}

\usepackage{PRIMEarxiv}
\usepackage{longtable}
\usepackage{array}
\usepackage{multirow}   
\usepackage{makecell}   
\usepackage{amsmath}
\usepackage{booktabs}
\usepackage{tabularx}
\usepackage{algorithm}
\usepackage{algpseudocode}
\usepackage{float}
\usepackage{multirow}
\usepackage{siunitx}
\usepackage{hyperref}
\algrenewcommand\algorithmicrequire{\textbf{Input:}}
\algrenewcommand\algorithmicensure{\textbf{Output:}}

\usepackage[utf8]{inputenc} 
\usepackage[T1]{fontenc}    
\usepackage{hyperref}       
\usepackage{url}            
\usepackage{booktabs}       
\usepackage{amsfonts}       
\usepackage{nicefrac}       
\usepackage{microtype}      
\usepackage{lipsum}
\usepackage{fancyhdr}       
\usepackage{graphicx}       
\usepackage{titlesec}       
\usepackage{enumitem}       
\usepackage{placeins}       
\graphicspath{{media/}}     

\titlespacing*{\section}{0pt}{1.2ex plus 0.2ex minus 0.2ex}{0.6ex plus 0.1ex}
\titlespacing*{\subsection}{0pt}{1.0ex plus 0.2ex minus 0.2ex}{0.4ex plus 0.1ex}
\titlespacing*{\subsubsection}{0pt}{0.8ex plus 0.2ex minus 0.2ex}{0.3ex plus 0.1ex}

\setlist{itemsep=2pt, topsep=2pt, parsep=0pt, partopsep=0pt}

\title{FRIENDS GUI: A Graphical User Interface for Data Collection
and Visualization of Vaping Behavior from a Passive Vaping Monitor%
\thanks{Published in \emph{Journal of Open Research Software},
14: 24 (2026). DOI: \href{https://doi.org/10.5334/jors.613}{10.5334/jors.613}.
Please cite the published version.}}

\author{
  \textbf{Shehan Irteza Pranto}\\
  Department of Electrical and Computer Engineering\\
  The University of Alabama\\
  \texttt{spranto@crimson.ua.edu}\\
  \and
  \textbf{Brett Fassler}\\
  Department of Electrical and Computer Engineering\\
  The University of Alabama\\
  \texttt{brfassler@crimson.ua.edu}\\
  \and
  \textbf{Md Rafi Islam}\\
  Department of Electrical and Computer Engineering\\
  The University of Alabama\\
  \texttt{mislam24@crimson.ua.edu}\\
  \and
  \textbf{Ashley Schenkel}\\
  Department of Psychology\\
  University of Buffalo\\
  \texttt{aschenke@buffalo.edu}\\
  \and
  \textbf{Larry W. Hawk}\\
  Department of Psychology\\
  University of Buffalo\\
  \texttt{lhawk@buffalo.edu}\\
  \and
  \textbf{Edward Sazonov}\thanks{\textit{Corresponding author}}\\
  Department of Electrical and Computer Engineering\\
  The University of Alabama\\
  \texttt{esazonov@eng.ua.edu}
}

\begin{document}
\maketitle

\begin{abstract}
Understanding puffing topography (PT), which includes puff duration, intra-puff interval, and puff count per session, is critical for evaluating Electronic Nicotine Delivery Systems (ENDS) use, toxicant exposure, and informing regulatory decisions. We developed FRIENDS (Flexible Robust Instrumentation of ENDS), an open-source device that can be attached to ENDS and records puffing and touching events. This paper introduces the FRIENDS graphical user interface (GUI) that improves accessibility and interpretability of data collected by FRIENDS. The GUI is a Python-based open-source tool that extracts, decodes, and visualizes 24-hour puffing data from the FRIENDS device. Validation using 24-hour experimental data confirmed accurate timestamp conversion, reliable event decoding, and effective behavioral visualization. The software is freely available on GitHub for public use.
\end{abstract}


\section{OVERVIEW}
\subsection*{INTRODUCTION}
The global use of Electronic Nicotine Delivery Systems (ENDS), commonly known as e-cigarettes or vapes, has increased significantly in recent years. In 2020, approximately 68 million people were using ENDS worldwide, nearly twice the estimated 35 million users in 2015 \cite{selamoglu2022general, jerzynski2021estimation}. ENDS use has grown substantially since 2010 in the United States, with 8.1 million adults identified as users by 2018 \cite{bradley2003two, nasem2018public, cobb2014fda}. As of 2021, 4.5\% of U.S. adults reported current ENDS use, with the highest prevalence (11\%) among young adults aged 18–24 \cite{kramarow2023current}. Although ENDS are often promoted as a safer alternative to combustible cigarettes, studies have shown that their vapor contains harmful chemicals, including acetaldehyde and formaldehyde, which are associated with respiratory and cardiovascular diseases \cite{ogunwale2017aldehyde}. Nevertheless, emerging evidence also suggests that ENDS can support smoking cessation in adults \cite{lindson2023pharmacological}.

A growing body of research has emphasized that  vaping patterns – such as the frequency, timing, and duration of puffs – play a critical role in smoking cessation outcomes \cite{harlow2022prospective, levy2018relationship, jackson2025trends}. However, a lack of standardized,  objective methods to measure vaping behavior hinders cross-study comparisons and limits understanding of usage trends \cite{doran2024protocol}. Most studies rely heavily on self-reporting, which is inherently limited \cite{chamberlain2013psychosocial, schuck2011effectiveness}. Many relevant behavioral metrics, such as puff timing and duration, cannot be accurately self-reported, and recall errors further reduce the reliability of such data \cite{shiffman2009immediate}. Even over a 4-hour session, self-reported puff counts correlated poorly with actual counts (r = 0.37) \cite{dowd2023examination}. Taking inspiration from sensors designed for monitoring of cigarette smoking \cite{lopez2013monitoring, sazonov2011rf}, sensors offer a promising alternative for more accurate monitoring of ENDS use. For example, a system described in \cite{adams2020puffpacket} enables detailed measurement of vaping behavior, including puff depth and intensity. However, its reliance on internal electrical connections limits its practical utility.

To address these challenges, the Flexible Robust Instrumentation of Electronic Nicotine Delivery Systems (FRIENDS) was developed as an open-source monitoring device that attaches externally to ENDS. FRIENDS integrates an electromagnetic (EM) sensor to detect puff events, a touch sensor to log when the device is held, and a thermistor to measure temperature changes during each puff and touch event. These components collectively record data to onboard memory with high-resolution ($15\,\mu\text{s}$) 64-bit extended POSIX timestamps with the event code.

This paper introduces the FRIENDS GUI, an open-source Python-based graphical user interface that supports seamless interaction with the FRIENDS device. The GUI facilitates data retrieval, processing, and conversion of raw binary logs into human-readable formats. It calculates key metrics such as event duration, enables users to set or synchronize device time, and supports flash memory management. The software also generates interactive, customizable visualizations that 
allow users to explore daily vaping patterns. Similar to other advanced processing tools \cite{soto2023optigui, evans2023hypat, spencer2023aquada, guo2023openseespyview}, the FRIENDS GUI aids in data-driven decision-making and supports refinement of experimental protocols.

\subsection*{IMPLEMENTATION AND ARCHITECTURE}
\begin{figure}[!t]
  \centering
  \includegraphics[width=\textwidth]{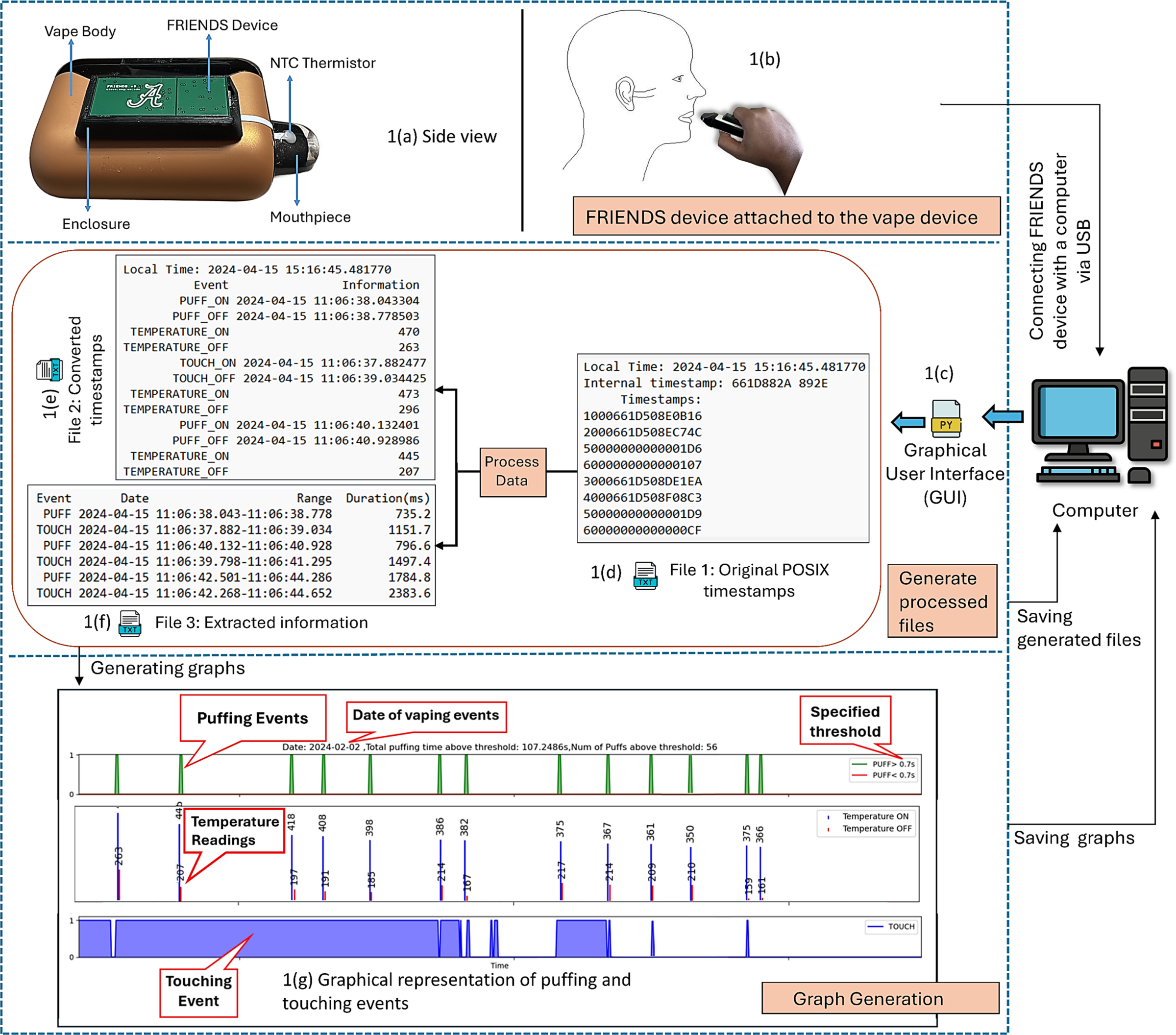}
  \caption{FRIENDS system architecture: (a) Device placement on ENDS; (b) user interaction with sensors; (c) graphical interface; (d) extended POSIX timestamp file; (e) converted timestamp file; (f) extracted event data; (g) graph of collected data, including EM detections, touches and temperature readings.}
  \label{fig:fig1}
\end{figure}

The FRIENDS system integrates hardware and software components to enable precise monitoring of vaping 
behavior. As shown in Figure \ref{fig:fig1}, the core hardware includes an MSP430G2433 microcontroller connected to three sensors: an electromagnetic (EM) sensor for detecting atomizer activity during puffs, a touch sensor for identifying when the device is held, and a negative thermal coefficient (NTC) thermistor for measuring temperature at the mouthpiece. The device attaches adhesively to ENDS and logs the start and end times of puff, touch events as well as aerosol temperature in internal flash memory, supporting continuous data collection for up to 10 days (Figure \ref{fig:fig1}a-b). The software components of the system have three primary responsibilities:

\begin{enumerate}
    \item \textbf{Data Collection Control:} The software facilitates synchronization of the device’s internal clock with the host computer and clears flash memory before a new session. This ensures accurate timestamping and reliable data capture for all events.
    
    \item \textbf{Data extraction and conversion:} The GUI extracts and decodes the binary data after data collection and provides key metrics, such as puff count, puff duration, and time of occurrence for a puff. Events are stored using extended POSIX timestamps (Figure \ref{fig:fig1}d), which include event codes, standard time, and fractional seconds (Table \ref{tab:table1}). Temperature is sampled at the start and end of puff and touch events, with values ranging from 0 to 1024, calculated via an internal voltage divider circuit. Although not timestamped independently, these readings are used to improve puff classification accuracy and eliminate false positives (Algorithm 1 in Appendix). Converted timestamps and extracted data, including puff durations and timing, are displayed in human-readable formats (Figure \ref{fig:fig1}e–f). 
    
    \item \textbf{Visualization:} The software generates 24-hour graphs showing puff, touch, and temperature data for the beginning and end of puffs. It also summarizes total puff duration and frequency per day (Figure \ref{fig:fig1}g).
\end{enumerate}

\begin{table}[htbp]
  \centering
  \caption{Formatting of original and converted timestamp}
  \label{tab:table1}
  \small
  \setlength{\tabcolsep}{4pt}
  \renewcommand\theadalign{tl}
  \renewcommand\theadfont{\bfseries\small}
  \begin{tabular}{@{}lllllp{3.5cm}@{}}
      \toprule
      \thead{EXTENDED POSIX \\ TIMESTAMP} 
      & \thead{EVENT \\ CODE} 
      & \thead{STANDARD \\ POSIX IN HEX} 
      & \thead{FRACTIONAL \\ PART} 
      & \thead{EVENT} 
      & \thead{CONVERTED \\ TIMESTAMP} \\
      \midrule
      100065CA42C88D44 & 1000 & 65CA42C8 & 8D44 & PUFF ON &
      PUFF\_ON 2024-02-12 10:09:44.551819 \\
      200065CA42C9AAE0 & 2000 & 65CA42C9 & AAE0 & PUFF OFF &
      PUFF\_OFF 2024-02-12 10:09:45.667480 \\
      50000000000001F9 & 5000 & -- & -- & TEMPERATURE ON &
      TEMPERATURE\_ON 505 \\
      6000000000000104 & 6000 & -- & -- & TEMPERATURE OFF &
      TEMPERATURE\_OFF 260 \\
      300065CA42CE04A0 & 3000 & 65CA42CE & 04A0 & TOUCH ON &
      TOUCH\_ON 2024-02-12 10:09:50.018066 \\
      400065CA42D007CD & 4000 & 65CA42D0 & 07CD & TOUCH OFF &
      TOUCH\_OFF 2024-02-12 10:09:52.030472 \\
      \bottomrule
  \end{tabular}
\end{table}

\begin{figure}[!t]
  \centering
  \includegraphics[width=1.0\textwidth]{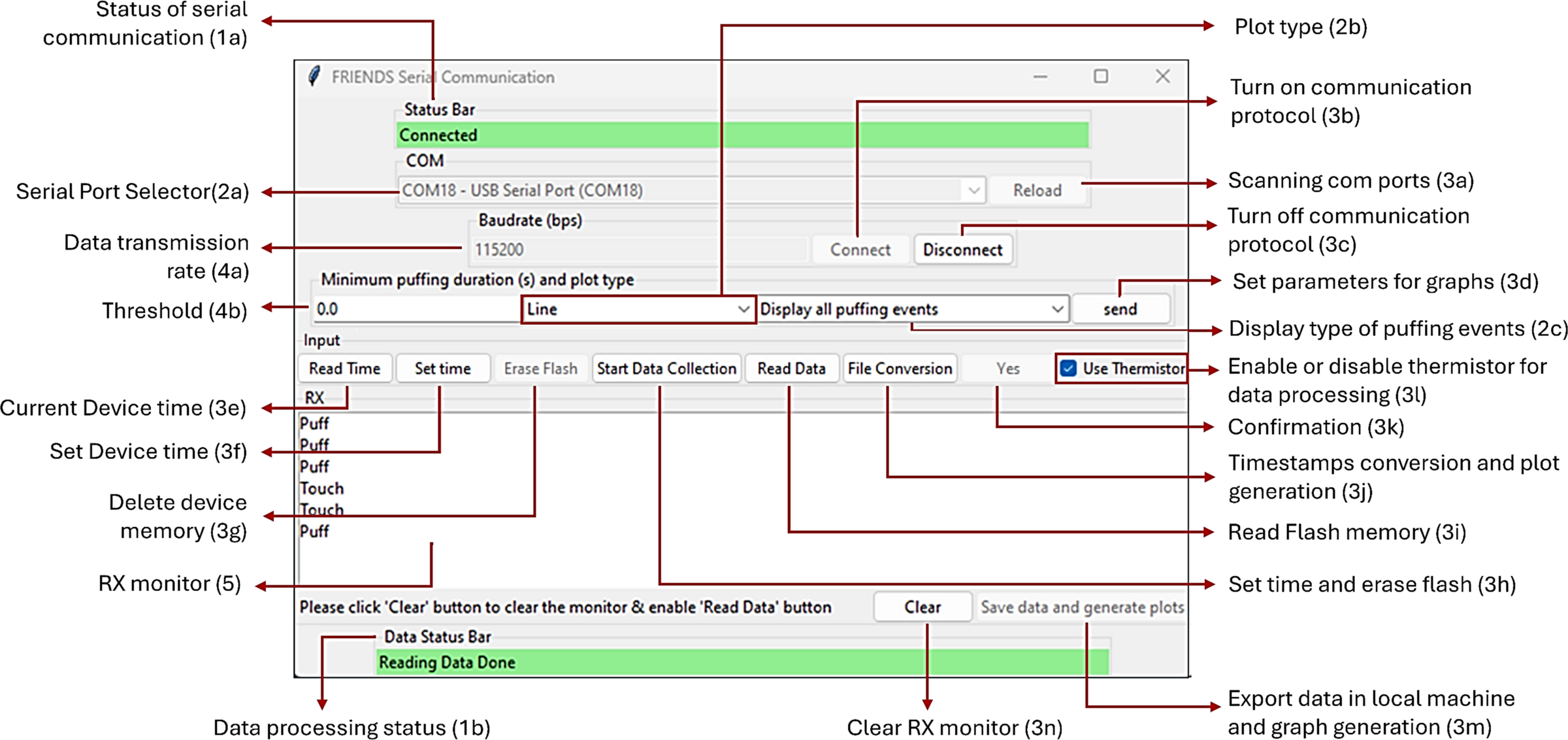}
  \caption{The main window of the FRIENDS GUI}
  \label{fig:fig2}
\end{figure}

\subsection*{GRAPHICAL USER INTERFACE}
User interaction with the FRIENDS system is facilitated through a graphical user interface (GUI), developed using Python 3.11 and the tkinter package. Multithreaded operations, implemented via Python’s threading library, allow real-time GUI responsiveness during serial communication. Data processing tasks such as timestamp conversion, event extraction, and formatting are handled using the Pandas and NumPy libraries, while Matplotlib is used for generating visualizations within the GUI. Figure \ref{fig:fig2} presents the GUI of the FRIENDS system, highlighting the key controls and data processing functions.

\begin{figure}[!t]
  \centering
  \includegraphics[width=\textwidth]{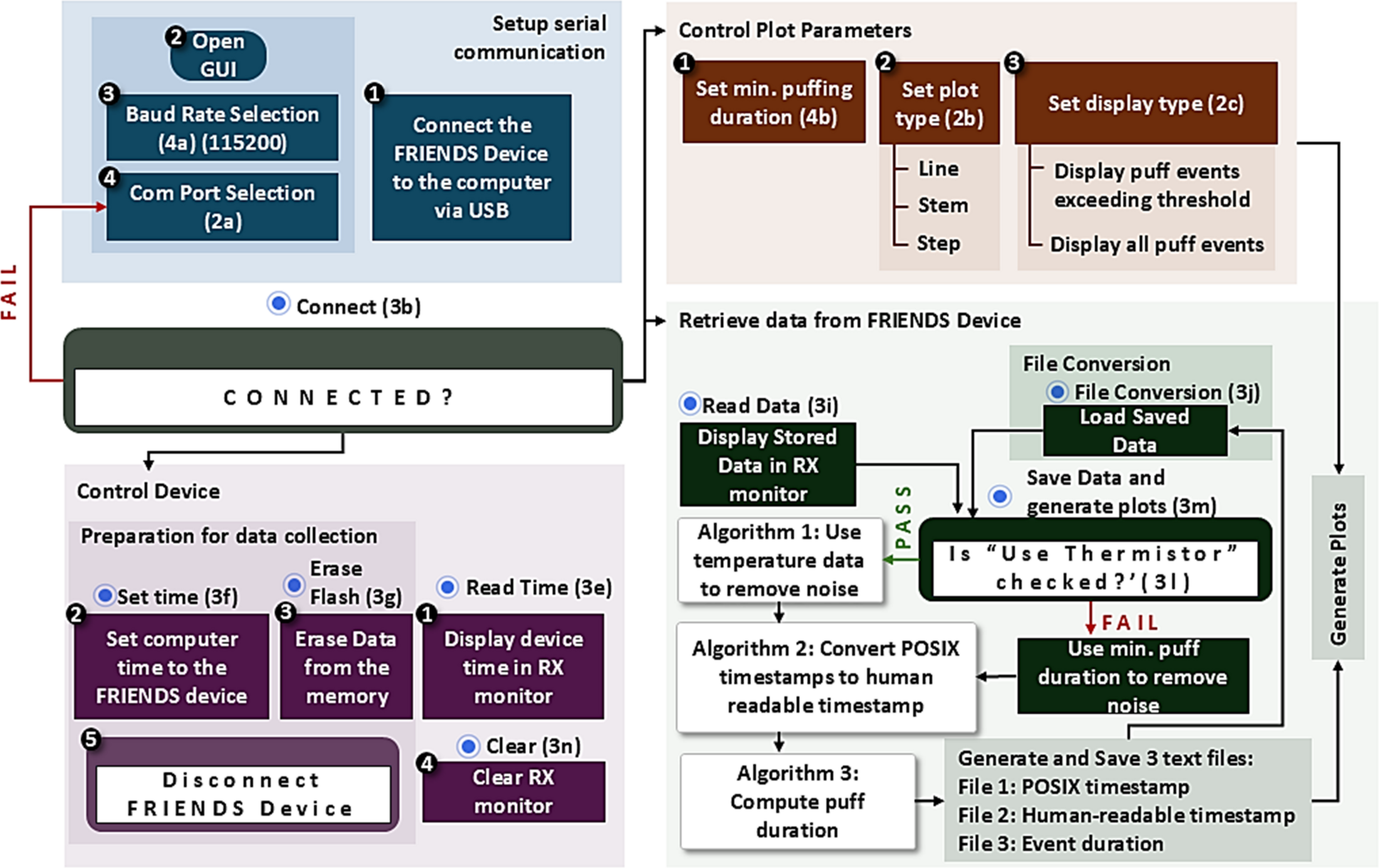}
  \caption{Flow Diagram of FRIENDS GUI}
  \label{fig:fig3}
\end{figure}

The GUI features five categories of widgets: (1) bars, (2) dropdown menus, (3) buttons, (4) entry fields, and (5) monitors, and the flow diagram in Figure \ref{fig:fig3} illustrates their functionalities.

Serial communication with the FRIENDS device is managed via the pyserial library. The “Connect” button establishes a link between the device and the host computer, and “Reload” refreshes the available COM port list.

Data collection is initialized by the “Start Data Collection” button, which clears the device’s flash 
memory and synchronizes its real-time clock with the host system. Users may also perform independent flash memory erasure (“Erase Flash”) or time synchronization (“Set Time”). The “Read Time” button displays the internal clock for verification purposes.

Data extraction begins by selecting “Read Data”, which initiates serial communication with the FRIENDS device. The device transmits 16-character hexadecimal records containing event type and timing, which are converted to extended POSIX timestamps (adjusted for local time), labelled by event type, and saved as a formatted text file (Figure \ref{fig:fig1}d).  

Data conversion runs automatically after extraction or can be triggered manually via the “File Conversion” button. Users may also access data using any open-source serial terminal emulator. Raw data from the FRIENDS device, saved in the text file, undergoes a multi-stage validation and cleaning process during pre-processing.

The software first checks for duplicate “Timestamps:” headers, which indicate that previously saved data may have been written again to the same file. If detected, a pop-up message alerts the user: \textit{“Multiple ‘Timestamps:’ headers detected: This indicates that the raw data was saved multiple times in the (input) text file. Please remove unwanted/unnecessary data from the text file.”} Leading spaces are then removed and the data are screened for corrupt entries that may occur during serial transmission, such as empty lines, incomplete records (fewer than 16 characters) and lines with more than 16 characters. Event logs are validated using logical pairing rules: puff events (codes 1/2), touch events (codes 3/4), and temperature events (codes 5/6) must each appear as valid on–off pairs in correct sequence. If errors are detected, the user is notified with diagnostic information containing two lists: \textit{“Line numbers with missing event pairs,”} which identifies lines with missing or incomplete event pairs, and \textit{“Line numbers with incomplete timestamps,”} which identifies lines with malformed or incomplete timestamps (e.g., fewer than 16 characters) in the raw data file. Unmatched or incorrectly ordered event pairs are removed. Additional consistency rules discard puff or touch pairs lacking a corresponding temperature event pair and remove duplicate temperature entries until a new event block begins. After pre-processing, the data conversion involves two steps: (1) translating extended POSIX timestamps to human-readable time, and (2) calculating event durations (Figure \ref{fig:fig1}d-f). 

The pseudo-code of processing data, converting timestamps, and extracting information is presented in algorithm 1, 2, and 3 in the Appendix, respectively.

Algorithm 1 processes extended POSIX timestamps using thermistor data when ``Use Thermistor'' is selected, improving puff detection for vape models where EM sensing is unreliable. Following testing across multiple vape models to determine cases in which puffs could be reliably detected using the FRIENDS device, a comprehensive library was developed that includes model-specific evaluations and identifies when thermistor data is required to improve puff detection performance \cite{claws_friends}. A temperature difference threshold of 10 is used to filter false positives. For example, a vape that has been idle for a long period may have an ADC value around 500; a change of 10, from 500 to 490, corresponds to a temperature increase from 25.98~$^\circ$C to 26.81~$^\circ$C (0.83~$^\circ$C). For a vape used recently, the baseline may be closer to 350; the same change of 10, from 350 to 340, corresponds to a temperature increase from 39.37~$^\circ$C to 40.37~$^\circ$C (1.00~$^\circ$C). If ``Use Thermistor'' is not selected, noise is removed based on a default minimum puff duration of 400ms. Algorithm 2 converts 64-bit extended POSIX timestamps into local time. It parses the event code, standard POSIX value, and fractional seconds, converting them into a readable datetime format. Algorithm 3 calculates the duration of events by identifying ON/OFF event pairs (e.g., PUFF\_ON/PUFF\_OFF) and computing their temporal difference. All results are stored in structured data frames for further visualization and analysis.

Visualization and graph generation in the FRIENDS GUI is initiated automatically upon selecting the “Save 
data and generate plots” or “File Conversion” button. The software first performs timestamp conversion and event 
extraction, then categorizes events by date and generates a separate plot (puff events at the top, temperature in 
the middle, touch at the bottom) for each day (Figure \ref{fig:fig1}g). Events are plotted on a 24-hour timeline (00:00 to 24:00). Users can customize plots by setting a minimum puff-duration threshold to display either all puff events or only those that meet the specified threshold.

\begin{table}[htbp]
  \centering
  \caption{Comparison of actual and software-analyzed data for puff quantity and duration}
  \label{tab:table2}
  \small
  \setlength{\tabcolsep}{4pt}
  \renewcommand\theadalign{tl}
  \renewcommand\theadfont{\bfseries\small}
  \begin{tabular}{@{}llllllll@{}}
      \toprule
       & \multicolumn{3}{l}{\thead{NUMBER OF PUFFS}} 
       & \multicolumn{3}{l}{\thead{TOTAL PUFFING DURATION (SEC)}} 
       & \thead{MISSING \\ DURATION} \\
      \cmidrule(lr){2-4} \cmidrule(lr){5-7}
       & \thead{ACTUAL}
       & \thead{RECORDED IN \\ DEVICE (RAW)}
       & \thead{RECORDED IN \\ DEVICE (1.0S \\ FILTER)}
       & \thead{ACTUAL}
       & \thead{RECORDED \\ IN DEVICE \\ (RAW)}
       & \thead{RECORDED \\ IN DEVICE \\ (1.0S FILTER)} 
       & \\
      \midrule
      Midnight   & 18 & 20 & 18 & 59.09 & 56.77 & 56.77 & 3.92\% \\
      Morning    & 18 & 19 & 18 & 57.85 & 59.06 & 59.06 & --2.09\% \\
      Afternoon  & 18 & 19 & 18 & 57.80 & 57.69 & 57.69 & 0.19\% \\
      Evening    & 18 & 31 & 18 & 57.27 & 58.54 & 57.24 & 0.05\% \\
      \midrule
      \textbf{Total}
                 & \textbf{72}
                 & \textbf{89}
                 & \textbf{72}
                 & \textbf{232.01}
                 & \textbf{232.06}
                 & \textbf{230.76} 
                 & \textbf{0.54\%} \\
      \bottomrule
  \end{tabular}
\end{table}

\begin{table}[htbp]
\centering
\caption{Assessment of Time Tracking, Conversion and Timing Discrepancies}
\label{tab:table3}
\footnotesize
\setlength{\tabcolsep}{3pt}
\renewcommand{\arraystretch}{1.0}
\begin{tabular}{@{}lllllllll@{}}
\toprule
\multirow{3}{*}{\textbf{SESSION}} &
\multirow{3}{*}{\shortstack[l]{\textbf{INTENDED}\\ \textbf{DURATION}\\ \textbf{(S)}}} &
\multirow{3}{*}{\shortstack[l]{\textbf{DURATION}\\ \textbf{(S) FROM}\\ \textbf{VAPE DISPLAY}}} &
\multicolumn{2}{l}{\multirow{2}{*}{\shortstack[l]{\textbf{FROM WEBSITE}\\ \textbf{(ACTUAL TIME)}}}} &
\multicolumn{2}{l}{\multirow{2}{*}{\shortstack[l]{\textbf{FROM GUI}\\ \textbf{(PROCESSED TIME)}}}} &
\multirow{3}{*}{\shortstack[l]{\textbf{APPROX.}\\ \textbf{AVERAGE TIME}\\ \textbf{DISCREPANCY}\\ \textbf{(a-a') OR (b-b')}}} \\
& & & \multicolumn{2}{l}{} & \multicolumn{2}{l}{} & \\
\cmidrule(lr){4-5} \cmidrule(lr){6-7}
& & & 
\shortstack[l]{\textbf{PUFF\_ON}\\ \textbf{TIME (a)}} & 
\shortstack[l]{\textbf{PUFF\_OFF}\\ \textbf{TIME (b)}} & 
\shortstack[l]{\textbf{PUFF\_ON}\\ \textbf{TIME (a')}} & 
\shortstack[l]{\textbf{PUFF\_OFF}\\ \textbf{TIME (b')}} & \\
\midrule
\multirow{3}{*}{Midnight}
& 2.0 & 2.28 & 01:37:02.13 & 01:37:04.50 & 01:37:01.18 & 01:37:03.37 & \multirow{3}{*}{1.0s} \\
& 3.0 & 3.19 & 01:40:31.96 & 01:40:35.13 & 01:40:30.61 & 01:40:34.11 & \\
& 4.0 & 4.34 & 01:44:37.26 & 01:44:41.60 & 01:44:36.98 & 01:44:40.36 & \\
\midrule
\multirow{3}{*}{Morning}
& 2.0 & 2.24 & 08:33:34.70 & 08:33:36.98 & 08:33:31.25 & 08:33:33.43 & \multirow{3}{*}{4.0s} \\
& 3.0 & 3.09 & 08:37:07.79 & 08:37:10.91 & 08:37:04.29 & 08:37:07.36 & \\
& 4.0 & 4.20 & 08:40:38.78 & 08:40:42.94 & 08:40:35.19 & 08:40:39.37 & \\
\midrule
\multirow{3}{*}{Afternoon}
& 2.0 & 2.21 & 16:47:01.76 & 16:47:04.01 & 16:46:56.18 & 16:46:58.34 & \multirow{3}{*}{6.0s} \\
& 3.0 & 3.12 & 16:50:05.20 & 16:50:08.29 & 16:49:59.43 & 16:50:02.52 & \\
& 4.0 & 4.12 & 16:54:05.75 & 16:54:10.16 & 16:54:00.28 & 16:54:04.42 & \\
\midrule
\multirow{3}{*}{Evening}
& 2.0 & 2.41 & 21:04:31.45 & 21:04:33.92 & 21:04:25.17 & 21:04:27.33 & \multirow{3}{*}{7.0s} \\
& 3.0 & 3.16 & 21:17:01.90 & 21:17:04.99 & 21:16:55.26 & 21:16:58.41 & \\
& 4.0 & 4.19 & 21:20:45.24 & 21:20:49.48 & 21:20:38.47 & 21:20:42.61 & \\
\bottomrule
\end{tabular}
\end{table}

\subsection*{QUALITY CONTROL}
The FRIENDS software was validated using a SMOK NORD 5 ENDS device, which has a puff-initiation button and displays puff duration. Validation experiments were conducted four times daily-at midnight, morning, afternoon, and evening. Each session consisted of three sets of six puffs, lasting approximately 2, 3, and 4 seconds respectively, totaling 18 puffs per session. A 30-second gap separated each puff. It is important to note that 
no human participants were involved in the validation experiments. Puffs were triggered either by a button 
pressed on the vape or a mechanical vaping machine \cite{fassler2025open}, and no human vaping intake occurred at any stage of the testing process. All experiments were recorded using the Zoom platform, with the camera directed at the NORD display and a shared window showing the real-time clock (\url{https://clock.zone/}) to log puff timing.

Actual puff timings (PUFF ON/OFF) and durations were extracted from the video recordings and compared with the data recorded by the FRIENDS device and processed by GUI. Table \ref{tab:table2} summarizes this comparison. The FRIENDS system recorded a total puff duration of 230.76 seconds—99.46\% of the actual 232.01 seconds and accurately detected all 72 puffing events. Seventeen false positives were captured, likely due to electromagnetic interference. The GUI’s filtering function, using a 1.0-second threshold, successfully identified and removed these events.
Table \ref{tab:table3} presents the precision of timestamp conversion, showing small timing drifts of up to 10 
seconds over a full day, which is acceptable for real-time monitoring.

\begin{figure}[!t]
  \centering
  \includegraphics[width=\textwidth]{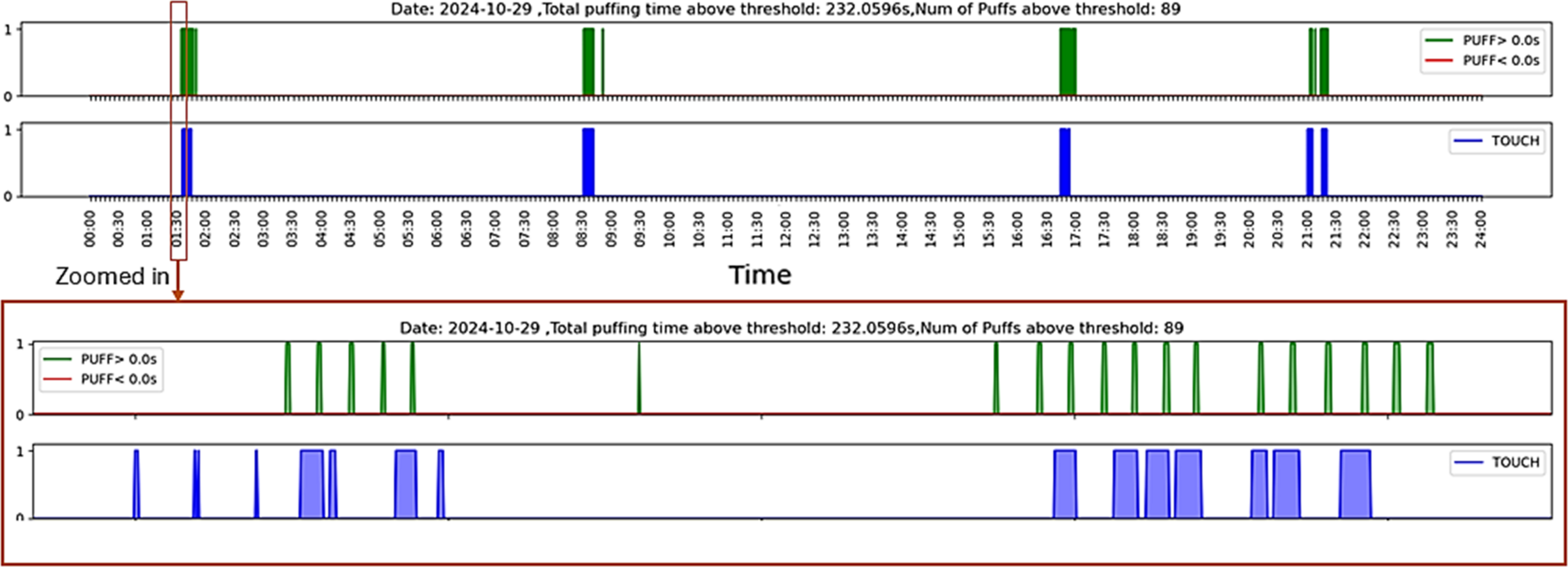}
  \caption{Graphical representation of puffing and touching events from the conducted experiment, recorded while the thermistor was turned off.}
  \label{fig:fig4}
\end{figure}

Figure \ref{fig:fig4} illustrates puffing and touching events over a 24-hour period, with a zoomed-in view near the 1:30-hour mark. While puffs generally occurred $\sim$30 seconds apart, a longer gap between puffs 5 and 6 and a likely false positive were observed.

This validation using 24-hour experimental data confirmed the software’s accuracy in timestamp conversion, data extraction, and graph generation, making it a practical tool for vaping behavior research across diverse study domains. 

To verify the software’s error-handling capability, file A (error-free) and file B (corrupted version of file A with intentionally introduced errors) were prepared. First, file A was processed to generate the reference output 
$A_{out}$, which contains timestamps converted to a human-readable format and the extracted event information. Corruptions introduced in file B include truncated timestamps, missing log lines, unexpected indentation, 
and stripped prefix bytes. The corrupted file B was then processed by the software to generate $B_{out}$. Finally, $B_{out}$ was compared with the reference output $A_{out}$ using Python’s filecmp.cmp() \cite{python_filecmp} function, which compares two files by checking their size and contents and returns True if they are identical or False if any difference is detected. The comparison successfully confirmed that all introduced corruptions were detected, verifying the correctness of the software. Subsequently, we tested the consistency between the GUI executable and standalone scripts by running both on file B and comparing their output text files using the same Python’s filecmp module. The comparison confirmed that both implementations produced identical results for the same input file. However, formal automated unit testing is acknowledged as a limitation of the current release and is planned for inclusion in future versions; in the current version, this gap is partially mitigated by comprehensive end-to-end validation of the full processing pipeline, which verifies the integrated performance of all key components under realistic usage conditions.

\section{AVAILABILITY}
\subsection*{OPERATING SYSTEM}
The software is compatible with any operating system that supports a standard Python installation, including the three major platforms: macOS, Windows, and Linux.

\subsection*{PROGRAMMING LANGUAGE}
Python 3.11 or higher.

\subsection*{ADDITIONAL SYSTEM REQUIREMENTS}
Input device: FRIENDS monitor

\subsection*{DEPENDENCIES}
Establishing device communication requires installation of FTDI VCP (Virtual COM Port) drivers on the computer where the FRIENDS GUI software is running. All necessary Python packages are specified in the requirements.txt file located in the root directory of the repository, allowing for reproducible setup. Key dependencies include Tkinter, NumPy, Matplotlib, Datetime, and Pyserial.
The FRIENDS GUI is standalone software built with Tkinter. Note that Tkinter availability depends on the operating system. The library is included with standard Python installations on Windows but may require separate installation on macOS systems, particularly when Python is installed via Homebrew (e.g., brew install python-tk). Linux users may similarly need to install it via their package manager (e.g., sudo apt-get install python3-tk on Debian/Ubuntu-based systems).

\subsection*{LIST OF CONTRIBUTORS}
\begin{enumerate}
    \item Shehan Irteza Pranto: Algorithm development, Software development, Experimental Design and Execution, Formal analysis, Writing (original draft)
    \item Brett Fassler: Software enhancement, Contribution to algorithm development, Writing (review and editing).
    \item Md Rafi Islam: Contribution to software testing and validation, Writing (review and editing)
    \item Ashley Schenkel: Project coordination
    \item Larry W. Hawk: Supervision, Writing (review and editing)
    \item Edward Sazonov: Conceptualization, Experimental Design, Supervision, Funding acquisition, Writing (review and editing)
\end{enumerate}

\subsection*{SOFTWARE LOCATION}

\textbf{Archive}\\
\hspace*{1.5em} \textit{Name:} Zenodo\\
\hspace*{1.5em} \textit{Identifier:} \href{https://doi.org/10.5281/zenodo.18753292}
{https://doi.org/10.5281/zenodo.18753292}\cite{pranto2024friendsgui}\\
\hspace*{1.5em} \textit{Licence:} MIT\\
\hspace*{1.5em} \textit{Publisher:} Shehan Irteza Pranto\\
\hspace*{1.5em} \textit{Date published:} 18/04/24

\textbf{Code repository}\\
\hspace*{1.5em} \textit{Name:} GitHub\\
\hspace*{1.5em} \textit{Identifier:} \href{https://github.com/eszsonic/FRIENDSGUI}{https://github.com/eszsonic/FRIENDSGUI}\\
\hspace*{1.5em} \textit{Licence:} MIT\\
\hspace*{1.5em} \textit{Date published:} 18/04/24

\subsection*{LANGUAGE}
English

\section{REUSE POTENTIAL}
While the FRIENDS GUI may not achieve the high volume of downloads obtained by consumer-facing software, its reuse potential within clinical and research communities is substantial. It fills a critical methodological gap by enabling objective, high-resolution measurement of real-world ENDS usage. The tool supports a variety of research applications, including our ongoing studies on individual differences in puffing patterns and their 
associations with nicotine exposure and dependence. These behavioral data are also expected to inform the 
development of improved self-report instruments. Moreover, FRIENDS and its GUI show strong potential for 
monitoring behavioral transitions, such as shifts in vaping frequency during smoking cessation or in response to 
public health interventions, areas of growing importance in nicotine research. Developed in Python, the GUI 
allows researchers to extract, decode, and visualize data collected by the FRIENDS device, including metrics such 
as total daily puff duration, individual puff length, puff count, and timing.

A key advantage of the FRIENDS GUI is its adaptability. The FRIENDS GUI can be employed for behavioral 
monitoring of tetrahydrocannabinol (THC) vapes or other electronic non-nicotine delivery systems (ENNDS) use. 
Furthermore, researchers can easily modify the source code to accommodate different ENDS detection devices or extend it for use with other custom-built behavioral monitoring systems. This flexibility makes it highly 
applicable across a wide range of research studies. In addition, the GUI requires no advanced programming 
skills for basic use, thanks to its intuitive interface and built-in functions for data conversion and visualization. This is important for use of the GUI by a broad range of nicotine and tobacco scholars and clinicians. The software is freely accessible under the MIT license, and hosted on GitHub (\url{https://github.com/eszsonic/FRIENDSGUI}), where users can also find supporting documentation. Its integration into existing research workflows streamline the process of data collection, extraction, and analysis, 
thereby accelerating behavioral research and increasing reproducibility. 

Validated through controlled experimental studies and already deployed in ongoing research on ENDS usage, the FRIENDS GUI holds strong potential for future clinical integration and the development of digital therapeutic tools. Researchers and clinicians are encouraged to adapt, expand, and contribute to its ongoing development. User support is provided through the GitHub issue tracker, where users can report bugs, request features, and ask questions. The authors are committed to providing prompt assistance. Comprehensive documentation is available in the repository, including the FRIENDS GUI User Guide (step-by-step instructions for setting up serial communication and installing the FRIENDS Serial Monitor), and the Installation Manual (instructions for building and packaging the FRIENDS software installer). In addition, the software is actively maintained, and users may contact the corresponding author directly via email for any queries regarding update releases or further support.

\appendix
\clearpage
\noindent\textbf{\Large APPENDIX}
\vspace{6pt}
 
\newlength{\algolinecolwidth}
\setlength{\algolinecolwidth}{2.2em}
 
\newcommand{\algoindent}{\hspace*{1.2em}}
 
\newcommand{\algoline}[2]{%
  \textit{#1.} & \textit{#2} \\
}
 
\newcommand{\algolinebf}[2]{%
  \textit{#1.} & \textbf{\textit{#2}} \\
}
 
\newcommand{\algoheader}[1]{%
  \multicolumn{2}{@{}l@{}}{\textbf{\textit{#1}}} \\
}
\newcommand{\algoinput}[1]{%
  \multicolumn{2}{@{}l@{}}{\textbf{\textit{Input:}} \textit{#1}} \\
}
\newcommand{\algooutput}[1]{%
  \multicolumn{2}{@{}l@{}}{\textbf{\textit{Output:}} \textit{#1}} \\
}
 
\newenvironment{algotable}[3]{%
  \par\noindent
  \begin{longtable}{@{}p{\algolinecolwidth}@{\hspace{0.6em}}|@{\hspace{0.6em}}p{0.85\linewidth}@{}}
  \hline
  \algoheader{#1}
  \hline
  \algoinput{#2}
  \algooutput{#3}
}{%
  \hline
  \end{longtable}
}
 
\begin{algotable}
  {Algorithm 1: Pseudocode for processing raw data}
  {A text file with extended POSIX timestamps \& Temperature Readings}
  {A dataframe with processed data and extended POSIX timestamps \& Temperature Readings}
\algoline{1}{Start}
\algoline{2}{DF = Dataframe with extended POSIX timestamp}
\algoline{3}{\textbf{\textit{If}} ``Use Thermistor'' Checkbox is selected, \textbf{\textit{then}}}
\algoline{4}{\algoindent DF2 = Initialize a dataframe with ``Timestamps'' column}
\algoline{5}{\algoindent \textbf{\textit{If}} `THERMISTOR\_ON' for Puff $N$+1 \textless= `THERMISTOR\_OFF' for Puff $N$, \textbf{\textit{then}}}
\algoline{6}{\algoindent\algoindent Concatenate puffs with `PUFF\_ON' from Puff $N$ and `PUFF\_OFF' from Puff $N$+1}
\algoline{7}{\algoindent\algoindent puff\_duration = `PUFF\_ON' -- `PUFF\_OFF'}
\algoline{8}{\algoindent\algoindent \textbf{\textit{If}} puff\_duration (ms) \textless= 400, \textbf{\textit{then}}}
\algoline{9}{\algoindent\algoindent\algoindent Read `THERMISTOR\_ON' and `THERMISTOR\_OFF'}
\algoline{10}{\algoindent\algoindent\algoindent temp\_diff = `THERMISTOR\_ON' -- `THERMISTOR\_OFF'}
\algoline{11}{\algoindent\algoindent\algoindent \textbf{\textit{If}} temp\_diff \textless= 10, \textbf{\textit{then}}}
\algoline{12}{\algoindent\algoindent\algoindent\algoindent Remove Puff}
\algoline{13}{\algoindent\algoindent\algoindent \textbf{\textit{Else if}} temp\_diff \textgreater\ 10, \textbf{\textit{then}}}
\algoline{14}{\algoindent\algoindent\algoindent\algoindent Keep Puff}
\algolinebf{15}{\algoindent\algoindent\algoindent end if}
\algolinebf{16}{\algoindent\algoindent end if}
\algolinebf{17}{\algoindent end if}
\algoline{18}{\textbf{\textit{Else if}} ``Use Thermistor'' Checkbox is not selected, \textbf{\textit{then}}}
\algoline{19}{\algoindent DF2 = Initialize a dataframe with ``Timestamps'' column}
\algoline{20}{\algoindent puff\_duration = `PUFF\_ON' -- `PUFF\_OFF'}
\algoline{21}{\algoindent \textbf{\textit{If}} puff\_duration (ms) \textless= 400, \textbf{\textit{then}}}
\algoline{22}{\algoindent\algoindent Remove Puff}
\algoline{23}{\algoindent \textbf{\textit{Else if}} puff\_duration (ms) \textgreater= 400, \textbf{\textit{then}}}
\algoline{24}{\algoindent\algoindent Keep Puff}
\algolinebf{25}{\algoindent\algoindent end if}
\algolinebf{26}{end if}
\algoline{27}{End}
\end{algotable}
 
\vspace{12pt}
 
\begin{algotable}
  {Algorithm 2: Pseudocode of generating converted timestamp}
  {A dataframe with extended POSIX timestamps}
  {Dataframe with local human readable time}
\algoline{1}{Start}
\algoline{2}{DF = Dataframe with extended POSIX timestamp}
\algoline{3}{DF2 = Initialize a dataframe with ``Timestamps'' column}
\algoline{4}{\textbf{\textit{for}} each row \textbf{\textit{in}} DF \textbf{\textit{do}}}
\algoline{5}{\algoindent From 64-bit timestamp, extract bits 16 to 47 (8 digits of hex) for POSIX time, and \newline \hspace*{1em} 48 to 63 (4 digits of hex) for Fraction of Seconds (FoS)}
\algoline{6}{\algoindent Convert 8 digits of hexadecimal POSIX (bits 16 to 47) to a decimal integer}
\algoline{7}{\algoindent Local time $\gets$ Convert  decimal integer to a local ``datetime'' object}
\algoline{8}{\algoindent FoS $\gets$ Convert 4 digits of hex value (bits 48 to 63) to decimal integer}
\algoline{9}{\algoindent Converted Time $\gets$ Append FoS to end of Local time}
\algoline{10}{\algoindent Event type $\gets$ Determine event type from bits 0 to 15 (4 digits of hex)}
\algoline{11}{\algoindent Add a new row in DF2 with Event type and Converted time}
\algolinebf{12}{end for}
\algoline{13}{End}
\end{algotable}
 
\vspace{18pt}
 
\begin{algotable}
  {Algorithm 3: Pseudocode for event information extraction}
  {Dataframe with converted timestamps}
  {Dataframe with extracted event information}
\algoline{1}{Start}
\algoline{2}{DF = Dataframe with converted timestamps}
\algoline{3}{DF2 = Initialize new dataframe with [`Event', `Date', `Range', `Duration (ms)'] columns}
\algoline{4}{Split Timestamp into ``Event'', ``Date'', and ``Time'' in DF}
\algoline{5}{\textbf{\textit{for}} each row \textbf{\textit{in}} ``Time'' column of DF:}
\algoline{6}{\algoindent Convert time into total seconds (with sub seconds) in a day}
\algolinebf{7}{end for}
\algoline{8}{\textbf{\textit{for}} each row \textbf{\textit{in}} DF:}
\algoline{9}{\algoindent \textbf{\textit{If}} current row is a `TOUCH\_ON' event followed by a `TOUCH\_OFF' event \textbf{\textit{then}}}
\algoline{10}{\algoindent\algoindent Event $\gets$ ``TOUCH''}
\algoline{11}{\algoindent\algoindent Date \ \ $\gets$ corresponding date from current row.}
\algoline{12}{\algoindent\algoindent Range $\gets$time of ``TOUCH\_ON'' event to time of ``TOUCH\_OFF'' event}
\algoline{13}{\algoindent\algoindent Duration (ms) $\gets$ (difference of seconds between ``TOUCH\_OFF'' and ``TOUCH\_ON'')*1000}
\algoline{14}{\algoindent\algoindent Add a new row in DF2 with  information of Event, Date, Range, and Duration (ms)}
\algolinebf{15}{\algoindent end if}
\algoline{16}{\algoindent \textbf{\textit{If}} current row is a `PUFF\_ON' event followed by a `PUFF\_OFF' event, \textbf{\textit{then}}}
\algoline{17}{\algoindent\algoindent Event $\gets$ ``PUFF''}
\algoline{18}{\algoindent\algoindent Date \ \ $\gets$ corresponding date from current row}
\algoline{19}{\algoindent\algoindent Range $\gets$ time of ``PUFF\_ON'' event to time of ``PUFF\_OFF'' event}
\algoline{20}{\algoindent\algoindent Duration (ms) $\gets$ (difference of seconds between ``PUFF\_OFF'' and ``PUFF\_ON'')*1000}
\algoline{21}{\algoindent\algoindent Add a new row in DF2 with information of Event, Date, Range, and Duration (ms)}
\algolinebf{22}{\algoindent end if}
\algolinebf{23}{end for}
\algoline{24}{End}
\end{algotable}

\subsection*{FUNDING STATEMENT}
This research was supported by the National Cancer Institute of the National Institutes of Health under Award Number R01CA266170. The content is solely the authors’ responsibility and does not necessarily reflect the official views of the National Institutes of Health.

\subsection*{COMPETING INTERESTS}
The authors have no competing interests to declare.

\subsection*{AUTHOR AFFILIATIONS}

\noindent\textbf{Shehan Irteza Pranto} \href{https://orcid.org/0000-0002-9818-4439}{orcid.org/0000-0002-9818-4439}\\
Graduate Research Assistant, Department of Electrical and Computer Engineering, The University of Alabama, Tuscaloosa, AL 35487, United States of America

\vspace{0.5em}
\noindent\textbf{Brett Fassler} \href{https://orcid.org/0009-0007-7510-2545}{orcid.org/0009-0007-7510-2545}\\
Graduate Research Assistant, Department of Electrical and Computer Engineering, The University of Alabama, Tuscaloosa, AL 35487, United States of America

\vspace{0.5em}
\noindent\textbf{Md Rafi Islam} \href{https://orcid.org/0000-0002-2180-5006}{orcid.org/0000-0002-2180-5006}\\
Graduate Research Assistant, Department of Electrical and Computer Engineering, The University of Alabama, Tuscaloosa, AL 35487, United States of America

\vspace{0.5em}
\noindent\textbf{Ashley Schenkel} \href{https://orcid.org/0009-0009-4569-8274}{orcid.org/0009-0009-4569-8274}\\
Clinical Research Coordinator, Department of Psychology, University of Buffalo, Buffalo, New York 14260-1660, United States of America

\vspace{0.5em}
\noindent\textbf{Larry W. Hawk} \href{https://orcid.org/0000-0001-6593-9746}{orcid.org/0000-0001-6593-9746}\\
Professor, Department of Psychology, University of Buffalo, Buffalo, New York 14260-1660, United States of America

\vspace{0.5em}
\noindent\textbf{Edward Sazonov} \href{https://orcid.org/0000-0001-7792-4234}{orcid.org/0000-0001-7792-4234}\\
Professor, Department of Electrical and Computer Engineering, The University of Alabama, Tuscaloosa, AL 35487, United States of America

\bibliographystyle{unsrt}  
\bibliography{references}

\end{document}